\documentclass[aps,prc,twocolumn,showpacs,floatfix]{revtex4-1}

\usepackage{graphicx}
\usepackage{dcolumn}
\usepackage{bm}
\usepackage{ulem}
\usepackage{color}

\newcommand{\al}{$\alpha$}

\newcommand{\SOB}{Sobiczewski}
\newcommand{\thisnuc}{$^{296}118$}
\newcommand{\lvii}{$^{292}$Lv}
\newcommand{\flviii}{$^{288}$Fl}
\newcommand{\cniv}{$^{284}$Cn}
\newcommand{\lvnull}{$^{290}$Lv}
\newcommand{\flvi}{$^{286}$Fl}
\newcommand{\cnii}{$^{282}$Cn}

\begin{document}

\title{
  $\alpha$-decay properties of $^{296}118$ from double-folding potentials
}

\author{Peter Mohr}
\email[Email: ]{WidmaierMohr@t-online.de; mohr@atomki.mta.hu}
\affiliation{
Diakonie-Klinikum, D-74523 Schw\"abisch Hall, Germany}
\affiliation{
Institute for Nuclear Research (Atomki), H-4001 Debrecen, Hungary}

\date{\today}

\begin{abstract}
$\alpha$-decay properties of the yet unknown nucleus $^{296}$118 are predicted
  using the systematic behavior of parameters of $\alpha$-nucleus
  double-folding potentials. The results are $Q_\alpha = 11.655 \pm
  0.095$\,MeV and $T_{1/2} = 0.825$\,ms with an uncertainty of about a factor
  of 4.
\end{abstract}


\maketitle

Very recently, \SOB\ \cite{Sob16} has analyzed the decay properties of the
yet unknown nucleus $^{296}118$ using a combination of $Q_\alpha$ values from
mass models and a phenomenological formula for the \al -decay half-lives. This
study was motivated by ongoing experiments which attempt to synthesize this
heaviest nucleus to date. The present work uses a completely different
approach which is based on the smooth and systematic bahavior of \al -decay 
parameters using double-folding potentials \cite{Mohr06}.

\SOB\ finds $Q_\alpha$ values between 10.93\,MeV and 13.33\,MeV from 9
different mass models. Using the phenomenological formula for \al -decay
half-lives of \cite{Par05}, the resulting half-lives for \thisnuc\ vary by
more than 5 orders of magnitude between 1.4\,$\mu$s and 0.21\,s. To reduce
this uncertainty, three mass models are identified in \cite{Sob16} which
describe the masses of nearby nuclei with the smallest deviations: Wang and
Liu (WS3+, \cite{Wang11}), Wang {\it et al.}\ (WS4+, \cite{Wang16,Wang14}),
and Muntian {\it et al.}\ (HN, \cite{Mun01,Sob07}). In detail, two \al -decay
chains are studied for this purpose: the known chain
$^{294}$118 $\rightarrow$ \lvnull\ $\rightarrow$ \flvi\ $\rightarrow$
\cnii\ (hereafter: ``chain-1''), 
and the chain
\thisnuc\ $\rightarrow$ \lvii\ $\rightarrow$ \flviii\ $\rightarrow$
\cniv\ (``chain-2'') where only the two latter \al -decays are known from
experiment. The selection of the mass formulae leads to a restricted range of
$Q_\alpha$ for \thisnuc\ from 11.62\,MeV (WS3+), 11.73\,MeV (WS4+), and
12.06\,MeV (HN), and the corresponding \al -decay half-lives are 4.8\,ms
(WS3+), 2.7\,ms (WS4+), and 0.50\,ms (HN). This range of predictions of almost
one order of magnitude for the \al -decay half-life of \thisnuc\ does not yet
include an additional uncertainty of the phenomenological formula of
\cite{Par05} which is on average a factor of 1.34 for even-even nuclei and
does not exceed a factor of 1.78 in most cases \cite{Par05}.

In a further study Budaca {\it et al.}\ \cite{Bud16} have applied empirical
fitting formulae for the prediction of the decay properties of \thisnuc . They
obtain a slightly lower $Q_\alpha = 11.45$\,MeV and half-lives of about
3\,ms. A very low value of $Q_\alpha = 10.185$\,MeV is derived from mass
formulae in \cite{San16,San14}, leading to predicted half-lives up to minutes
for \thisnuc . Half-lives of the order of 1\,ms have been obtained in
\cite{Shin16} using the WS4+ $Q_\alpha$ and various empirical formulae for the
half-life, and similar half-lives slightly below 1\,ms were found very
recently in \cite{Bao15,Zhang16} which are also based on $Q_\alpha$ from
WS4+.

For completeness it has to be mentioned that \al -decay is the dominant decay
mode of \thisnuc . Partial half-lives of \thisnuc\ for spontaneous fission
have been estimated in \cite{Sob16,San16b}; they exceed the \al -decay
half-life by several orders of magnitude.

Contrary to the study of \SOB\ and the other recent calculations for
\thisnuc\ \cite{Bud16,San16,San14,Shin16}, the present approach does not use
mass models for the prediction of the unknown $Q_\alpha$ of \thisnuc\ which is
the most important quantity for the prediction of its half-life. Instead, the
smooth behavior of parameters is used which is obtained in calculations with
systematic double-folding potentials \cite{Mohr06}. This method is
particularly well suited for the present case where the available experimental
results for chain-1 and chain-2 have to be extrapolated only to a very close
neighbor. For completeness it should be noted that there is another method
for an independent determination of $Q_\alpha$ from the systematics of $Q_\alpha$
differences of neighboring nuclei; unfortunately, the published values end at
$^{295}118$ and do not include \thisnuc\ \cite{Bao14}.

The application of double-folding potentials for \al -decay in a simple \al
+nucleus two-body model has been described in detail already in \cite{Mohr06},
and it has been applied and further developed in a series of \al -decay
studies in the last years (e.g.,
\cite{Kel16,Qian16,Ism16,Ism16b,Ni16,Ni15,Adel15,Ism15,Qian14,Qian14b,Ism13}).
Here I briefly repeat the essential points. First, the interaction between the
daughter nucleus and the \al -particle is calculated by a double-folding
procedure using an effective nucleon-nucleon interaction; for details, see
\cite{Mohr13}. As in \cite{Mohr06}, the unknown density of the daughter
nucleus is calculated from a 2-parameter Fermi distribution with the radius
parameter $R = R_0 A_D^{1/3}$ which scales with the mass number $A_D$ of the
daughter, and $R_0$ and the diffuseness $a$ are taken from the average values
of $^{232}$Th and $^{238}$U \cite{Vri87}. The density of the \al -particle is
also derived from from the charge density in \cite{Vri87}. This results in the
double-folding potential $V_{\rm{DF}}(r)$. The total potential is given by
\begin{equation}
V(r) = \lambda \, V_{\rm{DF}}(r) + V_{\rm{C}}(r)
\label{eq:pot}
\end{equation}
with the strength parameter $\lambda \approx 1.1 - 1.3$ for heavy nuclei
\cite{Mohr13,Atz96}. The Coulomb potential is calculated from the model of a
homogeneously charged sphere where the Coulomb radius $R_{\rm{C}}$ is taken
from the rooot-mean-square (rms) radius of the double-folding potential.

\begin{figure}
\includegraphics[ bb = 10 30 430 650, width = 85 mm, clip]{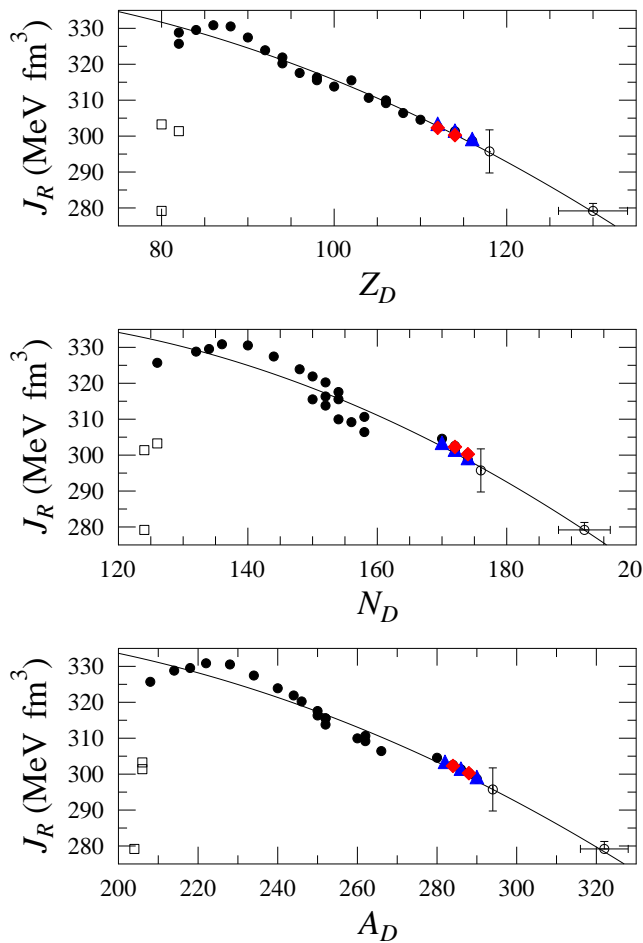}
\caption{
  \label{fig:jr2016} 
  (Color online)
  Volume integrals $J_R$ for superheavy nuclei as a function of $Z_D$
  (upper), $N_D$ (middle), and $A_D$ (lower). Data for chain-1 (blue
  triangles) and chain-2 (red diamonds) have been added. Otherwise, this
  figure is identical to Fig.~3 of my previous study \cite{Mohr06}; the lines
  are quadratic fits to the experimental data available in 2006.
}
\end{figure}

The strength parameter $\lambda$ is adjusted to reproduce the experimental
$Q_\alpha$; i.e., the potential $V(r)$ has an eigenstate at the correct energy
with a chosen number of nodes in the corresponding wave function ($N = 11$ in
the present case of $0^+$ ground states of even-even superheavy nuclei; see
\cite{Mohr06}). The resulting $\lambda$ values and volume integrals $J_R$ of
the nuclear potential are given in Table \ref{tab:para} for chain-1 and
chain-2. In addition, Fig.~\ref{fig:jr2016} shows $J_R$ as a function of the
proton number $Z_D$, neutron number $N_D$, and mass number $A_D$ of the
daughter nucleus. Fig.~\ref{fig:jr2016} is a copy of Fig.~3 of my previous
study \cite{Mohr06} where recent experimental data for chain-1 and chain-2
have been added. It is obvious from Fig.~\ref{fig:jr2016} that the volume
integrals $J_R$ show a regular and smooth dependence of $Z_D$, $N_D$, and
$A_D$, which can be used to obtain reliable estimates for unknown
nuclei. Discontinuities of $J_R$ appear only at shell closures, e.g.\ at the
doubly-magic daughter nucleus $^{208}$Pb (see Fig.~\ref{fig:jr2016} and
\cite{Mohr06}).
\begin{table*}[htb]
\caption{
\label{tab:para}
Parameters of the \al -decays in chain-1 and chain-2. Experimental values are
taken from \cite{Oga15}.
}
\begin{tabular}{ccclcccl}
\hline
& decay
& \multicolumn{1}{c}{$Q_\alpha$ (MeV)}
& \multicolumn{1}{c}{$\lambda$}
& \multicolumn{1}{c}{$J_R$ (MeV\,fm$^3$)}
& \multicolumn{1}{c}{$T_{1/2}^{\rm{calc}}$ (s)}
& \multicolumn{1}{c}{$T_{1/2}^{\rm{exp}}$ (s)}
& \multicolumn{1}{c}{$P$}
 \\
\hline
chain-1 & \flvi\ $\rightarrow$ \cnii       & 10.35 & 1.1633 & 302.86
& $8.48 \times 10^{-3}$   & $2.0 \times 10^{-1}$   & 0.0424 \\
chain-1 & \lvnull\ $\rightarrow$ \flvi     & 11.00 & 1.1568 & 300.96
& $7.36 \times 10^{-4}$   & $8.3 \times 10^{-3}$   & 0.0887 \\
chain-1 & $^{294}118$ $\rightarrow$ \lvnull & 11.82 & 1.1486 & 298.63
& $3.27 \times 10^{-5}$   & $6.9 \times 10^{-4}$   & 0.0473 \\
chain-2 & \flviii\ $\rightarrow$ \cniv & 10.07 & 1.1615 & 302.29
& $4.70 \times 10^{-2}$   & $6.6 \times 10^{-1}$   & 0.0713 \\
chain-2 & \lvii\ $\rightarrow$ \flviii & 10.78 & 1.1545 & 300.26
& $2.51 \times 10^{-3}$   & $1.3 \times 10^{-2}$   & 0.1930 \\
chain-2 & \thisnuc\ $\rightarrow$ \lvii 
& $11.655 \pm 0.095$\footnote{calculated using $\lambda = 1.1458 \pm 0.0010$}
& 1.1458\footnote{extrapolated from neighboring nuclei; see
  Fig.~\ref{fig:lambda}} & 297.80
& $7.30 \times 10^{-5}$
& $8.25 \times 10^{-4}$\footnote{$T_{1/2}^{\rm{predict}}$}   
& 0.0885\footnote{average of neighboring nuclei; see
  Fig.~\ref{fig:pre_av}} \\
\hline
\end{tabular}
\end{table*}

\begin{figure}
\includegraphics[ bb = 20 30 440 325, width = 85 mm, clip]{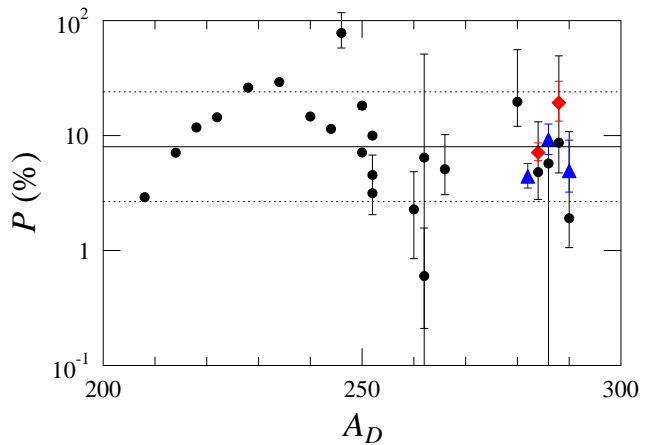}
\caption{
  \label{fig:pre2016} 
  (Color online)
  Preformation factors $P$ as a function of the mass number $A_D$ of the
  daughter nucleus, taken from \cite{Mohr06} and extended by data for chain-1
  (blue triangles) and chain-2 (red diamonds). The horizontal lines indicate an
  average value of $P \approx 8$\,\% (full line) and typical uncertainties of
  a factor of three (dotted lines); taken from \cite{Mohr06}.
}
\end{figure}

In a next step the \al -decay half-lives $T_{1/2,\alpha}^{{\rm{calc}}}$ are
calculated from the transmission through the barrier of the potential in
Eq.~(\ref{eq:pot}) using the semi-classical formalism of \cite{Gur87}. And
finally the preformation factor $P$ is calculated from the ratio
\begin{equation}
P = \frac{ T_{1/2,\alpha}^{\rm{calc}} }{ T_{1/2,\alpha}^{\rm{exp}} }
\label{eq:pre}
\end{equation}
The resulting preformation factors are shown in Fig.~\ref{fig:pre2016} which
is a repetition of Fig.~1 of \cite{Mohr06} with the additional results for
chain-1 and chain-2. An average value of about 8\,\% 
for P was found in \cite{Mohr06}, and the new data for chain-1 and chain-2 fit
nicely into this systematics.
Because \al -decay is the dominating decay mode of the nuclei in chain-1 and
chain-2 (except \flvi\ \cite{Oga15}), in the following the subscript $\alpha$
is omitted in $T_{1/2}$. 

\begin{figure}
\includegraphics[ bb = 20 30 430 320, width = 85 mm, clip]{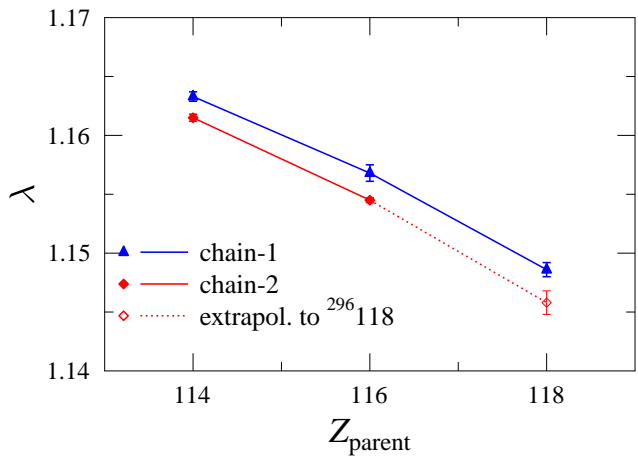}
\caption{
  \label{fig:lambda} 
  (Color online)
  Potential strength parameter $\lambda$ for chain-1 (blue triangles)
  and for chain-2 (red diamonds). The full symbols are derived from
  experimental data \cite{Oga15}; the open diamond is the extrapolation for
  the unknown nucleus \thisnuc . Further discussion see text.
}
\end{figure}

The very smooth and systematic behavior of the volume integrals $J_R$ in
Fig.~\ref{fig:jr2016} can be used for the prediction of unknown $Q_\alpha$
values. Instead of adjusting the strength parameter $\lambda$ to experimentally
known $Q_\alpha$, the strength parameter $\lambda$ is now fixed from neighboring
nuclei, and from the resulting potential $V(r)$ the eigenstate energy is
calculated. This is illustrated in Fig.~\ref{fig:lambda}: $\lambda = 1.1458
\pm 0.0010$ is estimated for \thisnuc . This estimate for $\lambda$ is well
constrained by the similar slope of $\lambda(Z)$ for chain-1 and chain-2 and
by the small and almost constant difference between chain-1 and chain-2.

The potential $V(r)$ with the strength parameter $\lambda = 1.1458$ has the
eigenstate with $N = 11$ nodes at $Q_\alpha = 11.655$\,MeV. The small
uncertainty of $\lambda$ translates to an uncertainty of $Q_\alpha$ of only
95\,keV. Thus, the present study predicts $Q_\alpha = 11.655 \pm 0.095$\,MeV
for the unknown nucleus \thisnuc . This result is very close to the
predictions of the selected mass models WS3+ and WS4+ and slightly lower than
the mass model HN \cite{Sob16}. It is interesting to note that already the
fits of $J_R$ in Fig.~\ref{fig:jr2016} (taken from \cite{Mohr06} and based on
the available data in 2006) predict $\lambda$ between 1.1413 and 1.1463 for
\thisnuc , corresponding to $Q_\alpha$ between 11.6\,MeV and 12.1\,MeV which
is almost exactly the range of $Q_\alpha$ from the three selected mass models
WS3+, WS4+, and HN in \cite{Sob16}.

Finally, the half-life of \thisnuc\ can be calculated from this potential with
$\lambda = 1.1458$. The result is $T_{1/2}^{\rm{calc}} =
73.0$\,$\mu$s. According to Eq.~(\ref{eq:pre}), for a prediction of the
experimental half-life $T_{1/2}^{\rm{exp}}$, the calculated half-life has to
be divided by the preformation factor $P$. Taking the average preformation
factor $P_{\rm{av}} = 0.0885$ of chain-1 and chain-2, one finally obtains
$T_{1/2}^{\rm{predict}} = 0.825$\,ms.

\begin{figure}
\includegraphics[ bb = 20 30 430 320, width = 85 mm, clip]{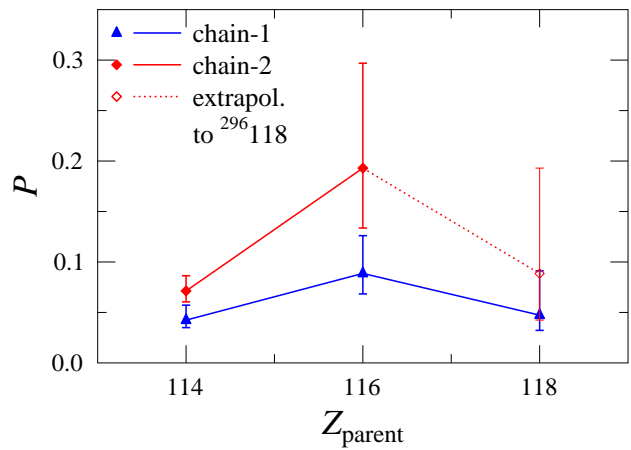}
\caption{
  \label{fig:pre_av} 
  (Color online)
  Extrapolation of the preformation factor $P$ to \thisnuc .
}
\end{figure}

A careful estimate of the uncertainty of the preformation factor $P$ can be
read from Fig.~\ref{fig:pre_av}. The average value of the 5 known $P$ in
chain-1 and chain-2 is $P_{\rm{av}} = 0.0885$. However, all $P$ have
significant uncertainties which result from the uncertainties of the
experimental \al -decay half-lives, and the $P$ vary between 0.0424 for
\flvi\ in chain-1 and 0.193 for \lvii\ in chain-2. Thus, I estimate the
uncertainty of $P$ for \thisnuc\ from the highest and smallest values of $P$
in chain-1 and chain-2, leading to $P = 0.0885^{+0.1045}_{-0.0461}$. Again it
is interesting to note that my earlier study in 2006 \cite{Mohr06} found very
similar values of $P \approx 0.08$ with an uncertainty of a factor of three.

The uncertainty of the predicted half-life $T_{1/2}^{\rm{predict}} =
0.825$\,ms can be estimated from the uncertainties of $Q_\alpha$ and $P$. The
uncertainty of $Q_\alpha$ of about 100\,keV translates to a factor of about
1.7 for the uncertainty of the half-life, and the uncertainty of $P$ of
slightly above a factor of two enters directly into the uncertainty of
$T_{1/2}^{\rm{predict}}$. Combining both uncertainties results in a factor of
about 4 uncertainty for the predicted half-life; i.e., the half-life of
\thisnuc\ should lie in between 0.2\,ms and 3.3\,ms.

In summary, I have used the smooth and regular behavior of the strength
parameter $\lambda$ of the \al -nucleus double-folding potential to estimate
the \al -decay energy $Q_\alpha$ of the unknown nucleus \thisnuc . The
prediction of $Q_\alpha = 11.655 \pm 0.095$\,MeV is completely independent of
mass formulae, but nevertheless in excellent agreement with the results from
the selected mass formulae in \cite{Sob16}. From the barrier transmission and
from the preformation $P$ of about 9\,\%, a half-life for \thisnuc\ of
0.825\,ms is predicted with an uncertainty of a factor of 4. These predictions
for the $Q_\alpha$ value and for the \al -decay half-life of \thisnuc\ may
help to guide experimentalists, and hopefully, these predictions can be
confronted with experimental results in the near future.

\acknowledgments 
I thank Zs.\ F\"ul\"op, Gy.\ Gy\"urky, G.\ G.\ Kiss, and E.\ Somorjai for many
encouraging discussions on \al -nucleus potentials. This work was supported by
OTKA (K108459 and K120666).


\begin{thebibliography}{99}
%
\bibitem{Sob16}
A.\ Sobiczewski,
\prc\ {\bf 94}, 051302(R) (2016).
%
\bibitem{Mohr06}
Peter Mohr,
\prc\ {\bf 73}, 031301(R) (2006); Erratum: \prc\ {\bf 74}, 069902(E) (2006).
%
\bibitem{Par05}
A.\ Parkhomenko and A.\ Sobiczewski,
Acta Phys.\ Pol.\ B {\bf 36}, 3095 (2005).
%
\bibitem{Wang11}
N.\ Wang and M.\ Liu,
\prc\ {\bf 84}, 051303(R) (2011).
%
\bibitem{Wang16}
Ning Wang, Min Liu, Xizhen Wu, Jie Meng,
\prc\ {\bf 93}, 014302 (2016).
%
\bibitem{Wang14}
N.\ Wang, M.\ Liu, X.\ Wu, J.\ Meng,
Phys.\ Lett.\ B {\bf 734}, 215 (2014).
%
\bibitem{Mun01}
I.\ Muntian, Z.\ Patyk, A.\ Sobiczewski,
Acta Phys.\ Pol.\ B {\bf 32}, 691 (2001).
%
\bibitem{Sob07}
A.\ Sobiczewski and K.\ Pomorski,
Prog.\ Part.\ Nucl.\ Phys.\ {\bf 58}, 292 (2007).
%
\bibitem{Bud16}
A.\ I.\ Budaca, R.\ Budaca, I.\ Silisteanu,
Nucl.\ Phys.\ A {\bf 951}, 60 (2016).
%
\bibitem{San16}
K.\ P.\ Santhosh, B.\ Priyanka, C.\ Nithya,
Nucl.\ Phys.\ {\bf A955}, 156 (2016).
%
\bibitem{San14}
K.\ P.\ Santhosh and B.\ Priyanka,
\prc\ {\bf 90}, 054614 (2014).
%
\bibitem{Shin16}
Eunkyoung Shin, Yeunhwan Lim, Chang Ho Hyun, Yogseok Oh,
\prc\ {\bf 94}, 024320 (2016).
%
\bibitem{Bao15}
X.\ J.\ Bao, S.\ Q.\ Guo, H.\ F.\ Zhang, Y.\ Z.\ Xing, J.\ M.\ Dong,
J.\ Q.\ Li,
J.\ Phys.\ G {\bf 42}, 085101 (2015).
%
\bibitem{Zhang16}
Shan Zhang, Yanli Zhang, Jianpo Cui, Yanzhao Wang,
\prc , accepted for publication.
%
\bibitem{San16b}
K.\ P.\ Santhosh and C.\ Nithya,
\prc\ {\bf 94}, 054621 (2016).
%
\bibitem{Bao14}
M.\ Bao, Z.\ He, Y.\ M.\ Zhao, A.\ Arima,
\prc\ {\bf 90}, 024314 (2014).
%
\bibitem{Kel16}
N.\ G.\ Kelkar and M.\ Nowakowski,
J.\ Phys.\ G {\bf 43}, 105102 (2016).
%
\bibitem{Qian16}
Yibin Qian and Zhongzhou Ren,
J.\ Phys.\ G {\bf 43}, 065102 (2016).
%
\bibitem{Ism16}
M.\ Ismail, A.\ Y.\ Ellithi, A.\ Adel, A.\ R.\ Abdulghany,
Nucl.\ Phys.\ {\bf A947}, 64 (2016).
%
\bibitem{Ism16b}
M.\ Ismail, A.\ Adel, M.\ M.\ Botros,
\prc\ {\bf 93}, 054618 (2016).
%
\bibitem{Ni16}
Dongdong Ni and Zhongzhou Ren,
\prc\ {\bf 93}, 054318 (2016).
%
\bibitem{Ni15}
Dongdong Ni and Zhongzhou Ren,
\prc\ {\bf 92}, 054322 (2015).
%
%
\bibitem{Adel15}
A.\ Adel and T.\ Alharbi,
\prc\ {\bf 92}, 014619 (2015).
%
\bibitem{Ism15}
M.\ Ismail, W.\ M.\ Seif, A.\ Y.\ Ellithi, A.\ Abdurrahman,
\prc\ {\bf 92}, 014311 (2015).
%
\bibitem{Qian14}
Yibin Qian and Zhongzhou Ren,
Phys.\ Lett.\ B {\bf 738}, 87 (2014).
%
\bibitem{Qian14b}
Yibin Qian and Zhongzhou Ren,
\prc\ {\bf 90}, 064308 (2014).
%
\bibitem{Ism13}
M.\ Ismail and A.\ Adel,
Nucl.\ Phys.\ {\bf A912}, 18 (2013).
%
\bibitem{Mohr13}
P.\ Mohr, G.\ G.\ Kiss, Zs.\ F\"ul\"op, D.\ Galaviz, Gy.\ Gy\"urky, 
E.\ Somorjai,
At.\ Data Nucl.\ Data Tables {\bf 99}, 651 (2013).
%
\bibitem{Vri87}
H.\ de Vries, C.\ W.\ de Jager, C.\ de Vries,
At.\ Data Nucl.\ Data Tables {\bf 36}, 495 (1987).
%
\bibitem{Atz96} 
U.\ Atzrott, P.\ Mohr, H.\ Abele, C.\ Hillenmayer, G.\ Staudt,
\prc\ {\bf 53}, 1336 (1996).
%
\bibitem{Gur87}
S.\ A.\ Gurvitz and G.\ K\"albermann,
\prl\ {\bf 59}, 262 (1987).
%
\bibitem{Oga15}
Yu.\ Ts.\ Oganessian and V.\ K.\ Utyonkov,
Nucl.\ Phys.\ {\bf A944}, 62 (2015).
%
\end{thebibliography}
\end{document}